\documentclass[letter,12pt]{article}
\usepackage{epsfig}
\usepackage{tabularx}
\usepackage{amsmath,amssymb}
\begin{document}
\setlength{\baselineskip}{25pt}
\renewcommand{\thefootnote}{\fnsymbol{footnote}}
\begin{flushright}
OSU-HEP-11-06
\end{flushright}
\begin{center}
{\Large Flavor Violation in a Minimal \\$SO(10)\times A_4$ SUSY GUT} \\
\vspace{1cm}
{\large Abdelhamid Albaid\footnote{\texttt{Email:abdelhamid.albaid@okstate.edu}}} \\
\vspace{0.3cm} {\it Department of Physics, Oklahoma State
University, Stillwater, OK, 74078 USA}

\end{center}
\begin{abstract}
Flavor violating processes in the quark and lepton sectors are
investigated within a realistic supersymmetric $SO(10)\times A_4$
grand unification model. By employing exotic heavy fermion fields,
this model successfully describes various features of the fermion
masses and mixings including large neutrino mixings accompanied by
small quark mixings. In this model the flavor violation is induced
at GUT scale, at which $A_4$ flavor symmetry is broken, as a
consequence of the large mixings of the light fermion fields with
these exotic heavy fields. The stringent experimental constraint
from $\mu\rightarrow e\gamma$ decay rate necessitates a high degree
of degeneracy of the supersymmetry breaking soft scalar masses of
the exotic heavy fields and supersymmetric scalar partners of the
light fermion fields. The choice of slepton masses of order 1 TeV is
found to be consistent with the constraints from branching ratio of
$\mu\rightarrow e\gamma$ and with all other flavor changing neutral
current processes being sufficiently suppressed.
\end{abstract}
\renewcommand{\thefootnote}{\arabic{footnote}}\setcounter{footnote}{0}

\section{Introduction}
Flavor changing neutral current (FCNC) processes impose severe
constraints on the soft supersymmetric breaking (SSB) sector of the
minimal supersymmetric standard model (MSSM). The simplest way to
satisfy the FCNC constraints is to adopt universality in the scalar
masses at a high energy scale where the effects of supersymmetry
(SUSY) breaking in the hidden sector is communicated to the scalar
masses of MSSM via gravitational interactions. For example, in the
the minimal supergravity model (mSUGRA) \cite{mSUGRA} the MSSM is a
valid symmetry between the weak scale and grand unification scale
($M_{\rm{GUT}}$) at which the universality conditions are assumed to
hold. In this case, the leptonic flavor violation (LFV) is not
induced. However, in a different class of models studied in Refs
\cite{MSSMRHN1, RGEs, MSSMRHN,SU(5)RHN,BHS, BPR} the universality of
the scalar masses will be broken by radiative corrections.
Consequently, FCNC will be induced in these models as discussed
below.

If the universality conditions hold at the grand unification scale
$M_{\rm{GUT}}$, the LFV is induced below GUT scale by radiative
corrections in the MSSM with right-handed neutrino
\cite{MSSMRHN1,RGEs, MSSMRHN} or SUSY-$SU(5)$ \cite{SU(5)RHN}
models. Unfortunately, it is difficult to predict LFV decay rates in
these models because the Dirac neutrino Yukawa couplings are
arbitrary within MSSM. However, in $SO(10)$ GUT model, we can
predict the LFV decay rates below the GUT scale because the Dirac
neutrino couplings are related to the up-type quark Yukawa couplings
and are thus fixed.

The FCNC could also be induced above the GUT scale by radiative
corrections. It was shown that as a consequence of the large top
Yukawa coupling at the unification scale, SUSY GUTs with
universality conditions valid at the scale $M^*$, where
$M_{\rm{GUT}}<M^*\leq M_{\rm{Planck}}$, predict lepton flavor
violating processes with observable rates \cite{BHS, BPR}. The
experimental search for these processes provides a significant test
for supersymmetric grand unification theory (SUSY GUT). Both
contributions of FCNC that are induced above and below
$M_{\rm{GUT}}$ will be studied in our model.

In this paper, the flavor violation processes for charged lepton and
quark sectors are investigated in the framework of a realistic SUSY
GUT model based on the gauge group $SO(10)$ and a discrete
non-abelian $A_4$ flavor symmetry \cite{albaid}. This model is
realistic because it successfully describes the fermion masses, CKM
mixings and neutrino mixing angles. This work differs from other
studies in several aspects. First, it is different from those based
on MSSM with right-handed neutrino masses or SUSY $SU(5)$ in the
sense that the Dirac neutrino Yukawa couplings are determined from
the fermion masses and mixing fit of the $SO(10)\times A_4$ model.
Thus, this model predicts the lepton flavor violation arising from
the renormalization group (RG) running from $M_{\rm{GUT}}$ to the
right-handed neutrino mass scales. Second, it is different from
those based on SUSY $SO(10)$ studied in \cite{SO(10)GUT} in the
sense that the FCNC processes are closely tied to fermion masses and
mixings. Finally, in the $SO(10)\times A_4$ model flavor violation
is induced at the GUT scale at which $A_4$ symmetry is broken due to
large (order one) mixing of the third generation of MSSM fields
($\psi_3$) with the exotic heavy fields ($\chi_i$, $i$ runs from 1
to 3). This large mixing arises when the $A_4$ flavor symmetry is
broken at the GUT scale. This is different from the case where the
flavor violation is induced due to large top Yukawa coupling at the
GUT scale \cite{BHS, BPR}. The reason for introducing the exotic
heavy fermion fields in our model is to obtain the correct fermion
mass relations at the GUT scale as we shall see in section 1. The
mass scales of these exotic fields range from $10^{14}$ GeV to
$10^{18}$ GeV depending on the values of the Yukawa couplings and
the scale of $A_4$ flavor symmetry breaking.

In this paper we study flavor violation of the hadronic and leptonic
processes by calculating the flavor violating scalar fermion mass
insertion parameters
$(\delta_{AB})_{ij}=\frac{(m^{2}_{AB})_{ij}}{\tilde{m}^2}$, for
$(A,B)=(L,R)$, with $\tilde{m}$ being the average mass of the
relevant scalar partner of standard model fermions (sfermions). All
the flavor violation sources are included in our calculations. The
sfermion mass insertions, $\delta_{LL,RR,LR}$, arise from the large
mixing between the $\psi_3$ and $\chi_i$ and the mass insertions,
$(\delta^{ij}_{LL})^{RHN}$, arise from RG running from
$M_{\rm{GUT}}$ to the right-handed neutrino mass scales. These
scalar mass insertion parameters are analyzed in the framework of
our model then they are compared with their experimental upper
bounds. We found that the most stringent constraint on flavor
violation comes from the $\mu\rightarrow e\gamma$ process. This
constraint requires a high degree of degeneracy of the soft masses
of MSSM fields and the exotic fields. Therefore, in this model we
assume that these soft masses are universal at the scale $M^*$ with
$M^*>M_{GUT}$, then we run them down to the GUT scale. The branching
ratio $Br(\mu\rightarrow e\gamma)$ close to experimental bound (i.e.
Br($\mu\rightarrow e\gamma$)=$1.2\times 10^{-11}$) is obtained when
the slepton masses of order $1$ TeV , while the Yukawa couplings
remain perturbative at the scale $M^*$. We also found in the
framework of our model that once the constraint from
$Br(\mu\rightarrow e\gamma)$ is satisfied, all the FCNC processes
will be automatically consistent with experiments.

This paper is organized as follows. In section 1, we show how the
fermion mass matrices are constructed in $SO(10)\times A_4$ model.
In section 2, we discuss the sources of flavor violation by finding
the sfermion mass insertion parameters $\delta^{ij}_{LL,RR}$ at the
GUT scale at which $A_4$ symmetry is assumed to be broken as well as
below the GUT scale. The results of the $SO(10)\times A_4$ model
regarding flavor violation analysis are presented in section 4.
Section 5 has our conclusion. The derivation of the light fermion
mass matrices and the light neutrino mass matrix after disentangling
the exotic fermions is shown in appendix A. In appendix B, we list
the renormalization group equations (RGEs) for various SUSY
preserving and breaking parameters between $M_{\rm{GUT}}$ and $M^*$
relevant for FCNC analysis.

\section{A Brief Review of Minimal $SO(10)\times A_4$ SUSY GUT}

In the $SO(10)$ gauge group, all the quarks and leptons of the SM
are naturally accommodated within a 16-dimensional irreducible
representation. However, minimal $SO(10)$ (i.e., with only one
10-dimensional Higgs representation) leads to fermion mass relations
at the GUT scale, such as $\frac{m^0_c}{m^0_t}=\frac{m^0_s}{m^0_b}$
and $m^0_{\mu}=m^0_s$, that are inconsistent with experiment. This
can be fixed by introducing exotic $16+\overline{16}$ fermions and
by coupling $16_i$ with these exotic fields via $45_H$, which is
used for $SO(10)$ symmetry breaking. The non-abelian discrete $A_4$
symmetry is chosen in our model because it is the smallest group
that has a 3-dimensional representation, so the three generations of
SM fields transform as triplet under $A_4$. Besides, FCNC is not
induced in the SUSY-$SO(10)\times A_4$  as long as $A_4$ symmetry is
preserved. However, as we will see later, the breaking of $A_4$
symmetry at the GUT scale will reintroduce the FCNC via large mixing
between the exotic and light fields. Based on the above reasons, a
$SO(10)\times A_4$ model is proposed in \cite{albaid}. In this
model, a minimal set of Higgs representations are used to break the
SO(10) gauge group to SM gauge group so the unified gauge coupling
remains perturbative all the way to the Planck scale. Employing this
minimal Higgs representation and $A_4$ symmetry, our model
successfully accommodates small mixings of the quark sector and
large mixings of the neutrino sector in the unified framework as
shown summarized  below.

The fermion mass matrices of the model proposed in \cite{albaid}
were constructed approximately. In this section, we construct these
matrices by doing the algebra exactly and show that the excellent
fit for fermion masses and mixings is obtained by slightly modifying
the numerical values of the input parameters of Ref.\cite{albaid}.
There are two superpotentials of the model. The first one
($W_{spin.}$) describes the couplings of the standard model fields
($\psi_i(16_i)$, $i$ runs from 1-3) with the exotic heavy
spinor-antispinor fields ($\chi_{i}(16_{i})$,
$\overline{\chi}_{i}(\overline{16}_{i})$, $i$ runs from 1 to 3),
while the second one ($W_{vect.}$) describes the couplings of
$\psi_i$ with the exotic 10-vector fields ($\phi_i$, $\phi'_i$,
$\phi''_i$, $\phi'''_i$, $i$ runs from 1 to 3) as given below:
\begin{eqnarray}
W_{spin.} &=& b_1 \psi_i \overline{\chi}_1 1_{Hi}+ b_2 \psi_i
\overline{\chi}_2 1'_{Hi} +
k_1 \chi_1 \overline{\chi}_3 45_H+ a \chi_3 \chi_2 10_H  + M_{\alpha} \chi_{\alpha} \overline{\chi}_{\alpha},\label{p32}\\
W_{vect.}&=& b_3\psi_i\phi_i16_H+M_{10}\phi_i\phi'_i+h'_{ijk}\phi'_i\phi'_j1_{Hk}+h_{ijk}\phi_i\phi_j1_{Hk}\nonumber\\
 &&+ A_{ijk} \phi'_i\phi''_j1''_{Hk}+m\phi''_i\phi'''_i+ k_2 \phi'''_i\phi'_i45_H\label{p33}.
\end{eqnarray}

The above superpotentials are invariant under $A_4$ and the
additional symmetry $Z_2\times Z_4\times Z_2$. The transformations
of the matter fields (i.e., the ordinary and exotic fermion fields)
and the Higgs fields under the assigned symmetry are given in Table
\ref{p3table1} and \ref{p3table2}.

\begin{table}
\center
\begin{tabular}{|c|c|c|c|c|c|c|} \hline
 SO(10)                     &$\psi_i$     &$\chi_1$,$\overline{\chi}_1$   &$\chi_2$,$\overline{\chi}_2$   &$\chi_3$,$\overline{\chi}_3$   &$Z^c_i$    \\\hline
 $A_4$                      & 3         & 1                         & 1                         & 1                         &3          \\\hline
 $Z_2\times Z_4\times Z_2$  &+,+,+      &+,-,+                      & -,+,+                     &+,+,-                      &+,+,+      \\\hline
 SO(10)                     &$\phi_i$     & $\phi'_i$                   &$\phi''_i$                   &$\phi'''_i$                  &$Z_i$      \\\hline
 $A_4$                      &3          &3                          &3                          &3                          &3          \\\hline
 $Z_2\times Z_4\times Z_2$  &+,$i$,+    &+,$-i$,+                   &+,$i$,-                    &+,$-i$,-                   &+,$-i$,+   \\\hline
\end{tabular}
\caption{The transformation of the matter fields under $SO(10)\times
A_4$ and $ Z_2 \times Z_4 \times Z_2 $.}\label{p3table1}
\end{table}

\begin{table}
\center
\begin{tabular}{|c|c|c|c|c|c|c|c|c|c} \hline
 SO(10)&$10_H$ & $45_H$ & $16_H$& $\overline{16}_H$&$1_{Hi}$ & $1'_{Hi}$&$1''_{Hi}$&$1'''_{Hi}$\\ \hline
 $A_4$ & 1 & 1 & 1 & 1&3&3&3&3\\   \hline
 $Z_2\times Z_4\times Z_2$ &-,+,-&+,-,-& +,$-i$,+ & +,$-i$,+&+,-,+&-,+,+&+,+,-&+,$i$,+\\   \hline
\end{tabular}
\caption{The transformation of the Higgs fields under
SO(10)$\times$$A_4$ and $ Z_2 \times Z_4 \times Z_2 $.}
\label{p3table2}
\end{table}

The general fermion mass matrix structure that results from
integrating out the exotic heavy spinor-antispinor fields in
$W_{spin.}$ is:
\begin{eqnarray}
M_F(spin.) =\left(\frac{aT_1T_2T_3 f^2\langle10_H\rangle
}{r_Fr_{F^c}}\right) \left(
\begin{array}{ccc}
0 & 0 & 0 \\
0 & 0 & Q_{F} s_{\theta}\frac{r_{F^c}}{f} \\
0 & Q_{F^c}s_{\theta}\frac{r_F}{f} & ( Q_F + Q_{F^c})c_{\theta}
\end{array} \right)\label{p312},
\end{eqnarray}
where we have made the following transformation:
$\psi_1\epsilon_1+\psi_2\epsilon_2+\psi_3\epsilon_3=\epsilon\psi'_3$
and
$\psi_1s_1+\psi_2s_2+\psi_3s_3=S(\psi'_2s_{\theta}+\psi'_3c_{\theta}$).
Here $\epsilon_i$ and $s_i$ are VEV-components of
$\langle1_H\rangle$ and $\langle1'_H\rangle$ respectively and
$s_{\theta}$($c_{\theta}$) is $\sin\theta$($\cos\theta$).
$f=(1+T^2_2+T^2_1(1+s^2_{\theta}T^2_2))^{-1/2}$
 and $r_F=(1+Q^2_FT^2_3T^2_1(1+s^2_{\theta}T^2_2)f^2)^{1/2}$ are
 factors that come from doing the algebra exactly (see appendix A). Here $T_1=\frac{b_1\epsilon}{M_1}$, $T_2=\frac{b_2S}{M_2}$,
$T_3=\frac{k_1\Omega}{M_3}$ and $Q=2I_{3R}+
\frac{6}{5}\delta(\frac{Y}{2})$ is the unbroken charge that results
from breaking $SO(10)$ to the SM gauge group by giving a VEV to
$45_H$, where $\left<45_H\right>=\Omega Q$. The charge Q for
different quarks and leptons is given as.
\begin{eqnarray}
Q_u=Q_d=\frac{1}{5}\delta, \hspace{0.5cm}  Q_{u^c}=-1-\frac{4}{5}\delta, \hspace{0.5cm} Q_{d^c}=1+\frac{2}{5}\delta, \nonumber \\
Q_{l}=Q_{\mu}=-\frac{3}{5}\delta, \hspace{0.5cm}
Q_{l^c}=1+\frac{6}{5}\delta, \hspace{0.5cm} Q_{\nu^c}=-1.
\end{eqnarray}

The above general structure of fermion mass matrix has the following
interesting features: (1) The relation $m^0_b=m^0_{\tau}$
automatically follows from $ Q_d + Q_{d^c}= Q_e + Q_{e^c}$, (2) The
hierarchy of the the second and third masses generation is obtained
by taking the limit $s_{\theta}\rightarrow 0$, and (3) The
approximate Georgi-Jarlskog relation $m^0_{\mu}=3m^0_s$ leads to two
possible values for $\delta$, either $\delta\rightarrow 0$ or
$\delta\rightarrow -1.25$, (4) the former possibility is excluded by
experiment since it leads to
$(m^0_c/m^0_t)/(m^0_s/m^0_{b})\rightarrow 1$ at the GUT scale, while
the latter possibility leads to
$(m^0_c/m^0_t)/(m^0_s/m^0_{b})\rightarrow 0$ which is closer to
experiments. Let us define $\delta=1+\alpha$. The masses and mixings
of the first generation arise from $W_{vector}$. The full mass
matrices arising from $W_{spinor}$ and $W_{vector}$ have the
following form:
\begin{eqnarray}
M_D &=& m^0_d\footnotesize\left(
\begin{array}{ccc}
0 & (c_{12}+\delta_3(\frac{3+2\alpha}{5}))r_dr_{d^c} & (-2\delta_2(\frac{3+2\alpha}{5})+\zeta )r_{d^c}\\
&&\\
(c_{12} & 0 & (2\delta_1(\frac{3+2\alpha}{5}) \\
-\delta_3(\frac{3+2\alpha}{5}))r_dr_{d^c} &&+s(\frac{-1+\alpha}{5})+\beta)r_{d^c}\\
&&\\
\zeta r_{d} &( s(\frac{3+2\alpha}{5})+\beta)r_{d} &
1\\
&&-2(\beta+\frac{3+2\alpha}{5}\delta_1)fc_{\theta}s_{\theta}T_2^2
\end{array} \right),\nonumber\\
M_U&=&m^0_u\footnotesize\left(
\begin{array}{ccc}
0 & 0 & 0 \\
0 & 0 & (\frac{1-\alpha}{5})s r_{u^c} \\0 & (\frac{1+4\alpha}{5})s
r_{u}& 1
\end{array} \right),\label{p31} \\
 M_L &=& m^0_d\footnotesize\left(
\begin{array}{ccc}
0 & (c_{12}+3\delta_3(\frac{-1+\alpha}{5}))r_er_{e^c} & (-\delta_2\alpha+\zeta)r_{e^c} \\
&&\\
(c_{12} & 0 & (\delta_1\alpha \\
-3\delta_3(\frac{-1+\alpha}{5}))r_er_{e^c}&&-3s(\frac{-1+\alpha}{5})+\beta)r_{e^c}\\
&&\\
 (\zeta &
(s(\frac{-1+6\alpha}{5})+\delta_1(\frac{6-\alpha}{5}) &
1\\
-\delta_2\frac{6-\alpha}{5})r_{e}&+\beta)r_{e}&-2(\beta+\frac{3+2\alpha}{5}\delta_1)fc_{\theta}s_{\theta}T_2^2
\end{array} \right),\nonumber\\
 M_N &=&m^0_u\footnotesize\left(
\begin{array}{ccc}
0 & 0 & 0 \\
0 & 0 & (\frac{-3+3\alpha}{5})s r_{\nu^c} \\0 & s r_{\nu} & 1
\end{array} \right)\nonumber,
\end{eqnarray}
where the parameters are defined in terms of the Yukawa couplings of
the superpotential ($W_{spin.}+W_{vect.}$) and the VEVs of the Higgs
fields as shown in appendix A. These matrices are multiplied by
left-handed fermions on the right and right-handed fermions on the
left. A doubly lopsided structure for the charged lepton and down
quark mass matrices of
 Eq.(\ref{p31}) can be obtained  by going to the limit
$\beta,\zeta,\alpha,\delta_3,c_{12},s\ll1$ and $ \delta_1,\delta_2$
are of order one. This doubly lopsided form leads simultaneously to
large neutrino mixing angles and to small quark mixing angles. Based
only on the above fermion mass matrices in Eq.(\ref{p31}), an
excellent fit is found for fermion masses (except for the neutrino
masses), quark mixing angles and neutrino mixing angles (except the
atmospheric angle) by giving the input parameters, appearing in
Eq.(\ref{p31}), the following numerical values: $\delta_1=-1.28$,
$\delta_2=1.01$, $\delta_3=0.015\times e^{4.95i}$, $\alpha=-0.0668$,
$ s=0.2897$, $\zeta=0.0126$, $ c_{12}=-0.0011e^{1.124i}$, and
$\beta=-0.11218$. The above numerical values lead to
$\sin\theta^L_{23}=0.92$ which is not close to the experimental
central value of atmospheric angle is $\sin\theta^{atm}_{23}=0.707$
\cite{f}. This contribution to the atmospheric angle is only from
the charged lepton sector. Therefore, the neutrino sector should be
included by considering the following superpotential:
\begin{eqnarray}
W_N=b_4\psi_iZ_i\overline{16}_H+h_{ijk}Z_iZ_j^c1'''_{Hk}+m_1Z^c_iZ^c_i\label{p34},
\end{eqnarray}
where two fermion singlets $Z_i$ and $Z^c_i$ that couple with the
Higgs singlet $1'''_{Hk}$ have been introduced.

The full neutrino mass matrix is constructed in Appendix B. The
Higgs singlet $1'''_{Hk}$ has the VEV-components ($\alpha_1$,
$\alpha_2$, $\alpha_3$). The light neutrino mass matrix is obtained
by employing the see-saw mechanism. The numerical values
($\alpha_1=0.075$, $\alpha_2=0.07$, $\alpha_3=0.9$, and
$\lambda=0.0465$ eV), where $\lambda$ is defined in appendix B, lead
to not only the correct contribution to the atmospheric angles
($\sin\theta^{atm}_{23}=0.811$) but also to the correct light
neutrino mass differences. The predictions of the fermion masses and
mixings are slightly altered by doing the algebra exactly compared
to the analysis of Ref.\cite{albaid}. These predictions and their
updated experimental values obtained from \cite{f} are shown in
Table \ref{d}. The right handed-neutrino masses arise from
integrating out the exotic fermion singlets $Z_i$ and $Z_i^c$ in
Eq.(\ref{p34}). The right handed-neutrino mass matrix is

\begin{eqnarray}
M_R=\Lambda\left(
\begin{array}{ccc}
\frac{\alpha^2_1}{\alpha^2_3} & \alpha_1\alpha_2(\frac{-1}{\alpha^2_3}+\frac{2}{\alpha^2_1+\alpha^2_2+\alpha^2_3})& \frac{-\alpha_1(\alpha^2_1-\alpha^2_2+\alpha^2_3)}{\alpha_3(\alpha^2_1+\alpha^2_2+\alpha^2_3)}\\
\alpha_1\alpha_2(\frac{-1}{\alpha^2_3}+\frac{2}{\alpha^2_1+\alpha^2_2+\alpha^2_3})& \frac{\alpha^2_2}{\alpha^2_3} & \frac{-\alpha_2(-\alpha^2_1+\alpha^2_2+\alpha^2_3)}{\alpha_3(\alpha^2_1+\alpha^2_2+\alpha^2_3)} \\
\frac{-\alpha_1(\alpha^2_1-\alpha^2_2+\alpha^2_3)}{\alpha_3(\alpha^2_1+\alpha^2_2+\alpha^2_3)}
&
\frac{-\alpha_2(-\alpha^2_1+\alpha^2_2+\alpha^2_3)}{\alpha_3(\alpha^2_1+\alpha^2_2+\alpha^2_3)}
& 1
\end{array} \right),
\end{eqnarray}
where $\Lambda=8.45\times10^{15}$ GeV and the right-handed neutrino
masses are given by $M_{R1}\approx M_{R2}\approx 1.4\times10^{12}$
GeV and $M_{R3}=8.5\times10^{15}$ GeV.

Another interesting feature of this model is that it contains a
minimal set of Higgs fields needed to break $SO(10)$ to the SM gauge
group. Consequently, the unified gauge coupling remains perturbative
all the way up to the Planck scale. This can be understood from the
running of the unified gauge coupling with energy scale
$\mu>M_{\rm{GUT}}$ as
\begin{eqnarray}
\frac{1}{\alpha}=\frac{1}{\alpha_G}-\frac{b_G}{2\pi}\log(\frac{\mu}{M_G}),
\end{eqnarray}
where $\alpha=g^2/(4\pi)$ and $b_G=S(R)-3C(G)$. Here $C(G)$ is the
quadratic Casimir invariant and $S(R)$ is the Dynkin index summed
over all chiral multiplets of the model. The unified gauge coupling
stays perturbative at the Planck scale (i.e $g(M_P)<\sqrt{2}$) as
long as $b_G<26$. Employing large Higgs representations might lead
to
 $b_G\geq 26$. For example, using $126_H$+$\overline{126}_H$ gives $b_G=46$. On the other hand, the $SO(10)\times A_4$-model gives
 $b_G=19$ which is consistent with the unified gauge coupling being
 perturbative till the Planck scale.
\begin{table}[t]
\center
\begin{tabular}{|c|c|c|c|c|c|c} \hline
 & Predictions & Expt. & Pull\\ \hline
 $m_c(m_c)$ & 1.4 & $1.27^{+0.07}_{-0.11}$ & 1.85 \\   \hline
 $m_t(m_t)$ & 172.5 & 171.3$\pm$2.3 & 0.52  \\   \hline
 $m_s/m_{d}$ & 19.4 & $19.5\pm2.5$ & 0.04 \\   \hline
 $m_s(2 Gev)$ & 109.6$\times 10^{-3}$ & $105^{+25}_{-35}\times 10^{-3} $ &  0.184 \\   \hline
 $m_b(m_b)$ & 4.31 & $4.2^{+0.17}_{-0.07}$ &  0.58  \\   \hline
 $V_{us}$ & 0.223 & 0.2255$\pm$0.0019 &  1.3 \\   \hline
 $V_{cb}$ & 38.9$\times 10^{-3} $ & (41.2$\pm$1.1)$\times 10^{-3} $ & 2 \\   \hline
 $V_{ub}$ & 4.00$\times 10^{-3} $ & (3.93$\pm$0.36)$\times 10^{-3} $& 0.7  \\   \hline
 $\eta$ & 0.319 & $0.349^{+0.015}_{-0.017}$ &1.7 \\\hline\hline
 $m_e(m_e)$ & 0.511$\times 10^{-3} $& 0.511$\times 10^{-3} $ & -  \\   \hline
 $m_{\mu}(m_\mu)$ & 105.6$\times 10^{-3} $ & 105.6$\times 10^{-3} $ &  - \\   \hline
 $m_{\tau}(m_\tau)$ & 1.776 & 1.776 & -  \\ \hline
 $\Delta m^2_{21}$ & $7.69\times10^{-3}\rm{eV}^2$  &$(7.59\pm0.2)\times10^{-3}\rm{eV}^2$  &0.5  \\   \hline
$\Delta m^2_{32}$ & $2.36\times10^{-3}\rm{eV}^2$
&$(2.43\pm0.13)\times10^{-3}\rm{eV}^2$  &0.5  \\   \hline
 $\sin\theta^{sol}_{12}$ & 0.555 & 0.566$\pm$0.018 & 0.61 \\   \hline
 $\sin\theta^l_{23}$ & 0.811 & 0.707$\pm$0.108 & 0.96 \\   \hline
 $\sin\theta_{13}$ & 0.141 & $<0.22$ & \\ \hline
\end{tabular}
\caption{The fermion masses and mixings and their experimental
values. The fermion masses, except the neutrino masses, are in GeV.
}\label{d}
\end{table}

We will use the same fit for fermion masses and mixings to calculate
the mass insertion parameters $\delta^{ij}_{LL,RR}$, and
$\delta^{ij}_{LR,RL}$ in the quark and lepton sectors and
consequently investigate the FCNC in this model. The charged lepton
and down quark mass matrices in Eq.(\ref{p31}) are diagonalized at
the GUT scale by bi-unitary transformation:
\begin{eqnarray}
M^{diag.}_{d,l}=V^{\dag d,l}_RM_{D,L}V^{d,l}_L,
\end{eqnarray}
where $V^{u,d,l}_{R,L}$ are known numerically. Now, we discuss the
sources of FCNC in this model.

\section{Sources of Flavor Violation in $SO(10)\times A_4$ Model}

We assume in our flavor violation analysis that $A_4$ flavor
symmetry is preserved above GUT scale and it is only broken at GUT
scale. In this case flavor violation is induced at GUT scale where
$A_4$ symmetry is broken. In this section we discuss the flavor
violation induced at the GUT scale by studying the sfermion mass
insertion parameter $\delta^{ij}_{LL,RR}$ and the chirality flipping
mass insertion ($A$-terms) parameter $\delta^{ij}_{LR,RL}$. We will
see that these flavor violation sources arise from large mixing of
the light fields with the heavy fields. This large mixing is due to
the breaking of $A_4$ symmetry. In addition, we discuss the induced
flavor violation arising below GUT scale through the RG running from
$M_{\rm{GUT}}$ to the right-handed neutrino mass scales.

\subsection{The Scalar Mass Insertion Parameters}
Let us assume the soft supersymmetry breaking terms originate at the
messenger scale $M^*$, where $M_{\rm{GUT}}<M^*\leq M_{\rm{Planck}}$.
The quadratic soft mass terms of the matter superfields that appear
in the superpotential $W_{spin.}$ are
\begin{eqnarray}
-\mathcal{L}=\tilde m^2_{\psi} \psi^{\dag}_i \psi_i +\tilde
m^2_{\chi_{i}}\chi^{\dag}_{i}\chi_{i}+\tilde
m^2_{\overline{\chi}_{i}}\overline{\chi}^{\dag}_{i}\overline{\chi}_{i}\label{p35}.
\end{eqnarray}

The MSSM scalar fermions that reside in $\psi_i$ transform as
triplets under the non-abelian $A_4$ symmetry. Since the $A_4$
symmetry is intact, they have common mass ($\tilde m^2_{\psi}$) at
the scale $M^*$. On the other hand, the exotic fields each of which
transforms as singlet under $A_4$ symmetry have different masses
($\tilde m^2_{{\chi}_{i}}$, $\tilde m^2_{\overline{\chi}_{i}}$, $i$
runs 1-3) at the scale $M^*$.

The MSSM scalars remain degenerate above the GUT scale where the
$A_4$ symmetry is broken. In order to find the scalar masses in the
fermion mass eigenstates, two transformations are required. The
first transformation is needed to block-diagonalize the fermion mass
matrix into a light and a heavy blocks as shown in Appendix A. The
upper left corner represents the $3\times3$ light fermions mass
matrix. The second transformation is the complete diagonalization of
the light fermion mass matrix. Applying the first transformation to
the quadratic soft mass terms of Eq.(\ref{p35}) by going to the new
orthogonal basis ($L_2$, $L_3$, $H_1$, $H_2$, $H_3$) as defined in
appendix A, the quadratic soft mass matrix of the light states is
transformed as follows:
\begin{eqnarray}
\tilde m^2_{{\psi}}I
 \rightarrow \tilde m^2_{{\psi}}I+\delta\tilde m^2_{{\psi}},
\end{eqnarray}
where,
\begin{eqnarray}
\delta \tilde m^2_{{\psi}}=\left(
\begin{array}{ccc}
0& 0 & 0\\
0  & 0& \epsilon  \\
0&\epsilon & \delta
\end{array} \right),\label{p3LL1}
\end{eqnarray}
$\epsilon=\frac{f}{r_F}T^2_2s_{\theta}(\tilde
m^2_{{\chi}_{2}}-\tilde m^2_{{\psi}})$,
$\delta=((\frac{f}{r_F})^2-1)\tilde
m^2_{{\psi}}+(\frac{f}{r_F})^2(\tilde m^2_{{\chi}_{1}}T^2_1+\tilde
m^2_{{\chi}_{2}}T^2_2+\tilde m^2_{\chi_{3}}Q^2T^2_1T^2_3)$, and we
have safely ignored the terms that contain $s_{\theta}^2\ll1$.  It
is obvious that the first two generations of the light scalars are
almost degenerate because the mixing of the second light generation
($L_2$) with the heavy states is proportional to $s_{\theta}\ll1$.
On the other hand, since the mixing of the third light generation
($L_3$) with the heavy states is of order one, its mass splits from
those of the first two generations.

The top Yukawa coupling is given in terms of $T_1$, $T_2$, and $T_3$
as:
\begin{eqnarray}
Y_t=\frac{af^2(Q_u+Q_{u^c})T_1T_2T_3}{r_{u^c}r_u}.
\end{eqnarray}
The numerical values of $T_1=0.0305$, $T_2=2$, $T_3=100$ and
$a\sim1.2$ are consistent with the top Yukawa coupling at the GUT
scale to be of order $\lambda^{GUT}_t\sim 0.5$ and $r_{u,u^c}$ to be
of order one. Plugging these numerical values and
$s_{\theta}=0.0465$ into the expressions for $\epsilon$ and $\delta$
gives us:
\begin{eqnarray}
(\delta_d, \delta_{d^c}, \delta_e, \delta_{e^c})&=&(0.81,0.87,0.88,0.82)(\tilde m^2_{\chi}-\tilde m^2_{\psi}) \nonumber, \\
(\epsilon_d, \epsilon_{d^c}, \epsilon_e,
\epsilon_{e^c})&=&(0.061,0.05,0.048,0.06)(\tilde m^2_{\chi}-\tilde
m^2_{\psi}).\label{p3LL2}
\end{eqnarray}
Here we have dropped $\tilde m^2_{\chi_1}$ terms because their
coefficients are negligible. Also, the RGE expressions of $\tilde
m^2_{\chi_2}$ and $\tilde m^2_{\chi_3}$ are the same (see
Eq.(\ref{p3r})), so we have assumed that $\tilde m^2_{\chi_2}=\tilde
m^2_{\chi_3}=\tilde m^2_{\chi}$.

The next step is to apply the second transformation by evaluating
$V^{\dag d,l}_L\delta m^2_{\psi}V^{d,l}_L$ and similarly for
$L\rightarrow R$. The unitary matrices $V^{d,l}_L$ are numerically
known from the fitting for fermion masses and mixings. So, the mass
insertion parameters for charged leptons and down quarks are given
respectively by
\begin{eqnarray}
(\delta^{d,e}_{LL,RR})_{ij}&=&(V^{\dag d,l}_{L,R}\delta
\tilde{m}^2_{d,l}V^{d,l}_{L,R})_{ij}/\tilde{m}^2_{d,l}.
\label{p3LL3}
\end{eqnarray}
The above mass insertion analysis without including the
superpotential $W_{vect.}$ is good enough because we assumed in our
analysis that the mixing of the $10$ vector multiplets with the
ordinary spinor fields is small.
\subsection{The Chirality
Flipping Mass Insertion ($A$-terms)}

The FV processes are also induced from the off-diagonal entries of
the chirality flipping mass matrix $\tilde M_{RL}$. The chirality
flipping soft terms are divided into two parts $\mathcal{L}_{spin}$
and $\mathcal{L}_{vect}$:
\begin{eqnarray}
-\mathcal{L}_{spin}&=& \tilde b_1 b_1 \tilde{\psi}_i
\tilde{\overline{\chi}}_1 1_{Hi}+ \tilde b_2 b_2 \tilde{\psi}_i
\tilde{\overline{\chi}}_2 1'_{Hi} + \tilde k_1 k_1\tilde{\chi}_1
\tilde{\overline{\chi}}_3 45_H\nonumber\\
&&+\tilde a a\tilde{\chi}_3 \tilde{\chi}_2 10_H  + \tilde G_{i}M_i\tilde{\chi}_{i} \tilde{\overline{\chi}}_{i},\label{p36}\\
-\mathcal{L}_{vect}&=& \tilde b_3 b_3 \tilde{\psi}_i\tilde{\phi}_i16_H+\tilde{B}_{10}M_10\tilde{\phi}_i\tilde{\phi'}_i+\tilde{h'}_{ijk}h'_{ijk}\tilde{\phi'}_i\tilde{\phi'}_j1_{Hk}+\tilde{h}_{ijk}h_{ijk}\tilde{\phi}_i\tilde{\phi}_j1_{Hk}\nonumber\\
 &&+ \tilde{A}_{ijk}A_{ijk} \tilde{\phi'}_i\tilde{\phi''}_j1''_{Hk}+\tilde{g}m\tilde{\phi''}_i\tilde{\phi'''}_i+ \tilde{k}_2k_2 \tilde{\phi'''}_i\tilde{\phi'}_i45_H.\label{p3soft}
\end{eqnarray}

The fourth term of Eq.(\ref{p36}) induces the off-diagonal elements
of the chirality flipping mass matrix, if it is written in terms of
the new orthogonal basis defined in Eqs.(\ref{p3bc1}). This
transformation can be represented by
\begin{eqnarray}
\tilde M_{RL}^2(spin.)\rightarrow \tilde a M_{F}(spin.),
\end{eqnarray}
where $M_F(spin.)$ is defined in Eq.(\ref{p312}). The entire
chirality flipping mass matrix in the new orthogonal basis is
obtained by including $-\mathcal{L}_{vect}$. The  bi-unitary
transformations that block-diagonalize the full fermion mass matrix
is applied on the entire chirality flipping mass matrix (see
Appendix A). Accordingly, the $3\times3$ quadratic mass matrix
($\tilde{M}_{LR}^2$) associated with the light states is transformed
as follows:
\begin{eqnarray}
\tilde M_{RL}^2\rightarrow \tilde a M_{F}(spin.)+\tilde b_3
M_{F}(vector),
\end{eqnarray}
where $M_{F}(vect.)= -mM^{-1}M'$ (see Eq.(\ref{p310})) and we have
assumed for simplicity that the soft parameters appearing in
Eq.(\ref{p3soft}) are all of the same order. Then, the $M_{LR}^2$
matrix is written in the fermion mass eigenstate basis as:
\begin{eqnarray}
\tilde M_{RL}^2\rightarrow V^{\dag}_{R}(\tilde a M_{F}(spin.)+\tilde
b_3 M_{F}(vect.))V_{L}.
\end{eqnarray}
It is straightforward to show that the chirality  mass insertion
parameters are given by:
\begin{eqnarray}
(\delta_{RL})_{ij}= \frac{\tilde b_3}{\tilde m^2_f}
M_{Fi}^{diag.}\delta_{ij}+ (\tilde z V^{\dag}_{R}
M_{F}(spinor)V_{L})_{ij}\label{p3d}, \label{p3LR}
\end{eqnarray}
where $M_F^{diag.}=V^{\dag}_{R} M_{F}V_{L}$ and $\tilde
z=\frac{\tilde a-\tilde b3}{\tilde m^2_f}$. The induced FV arises
only from the second term of Eq.(\ref{p3d}).

\subsection{Mass Insertion Parameters Induced Below $M_{\rm{GUT}}$}
\begin{figure}[t]
\centering {\includegraphics[width=10 cm]{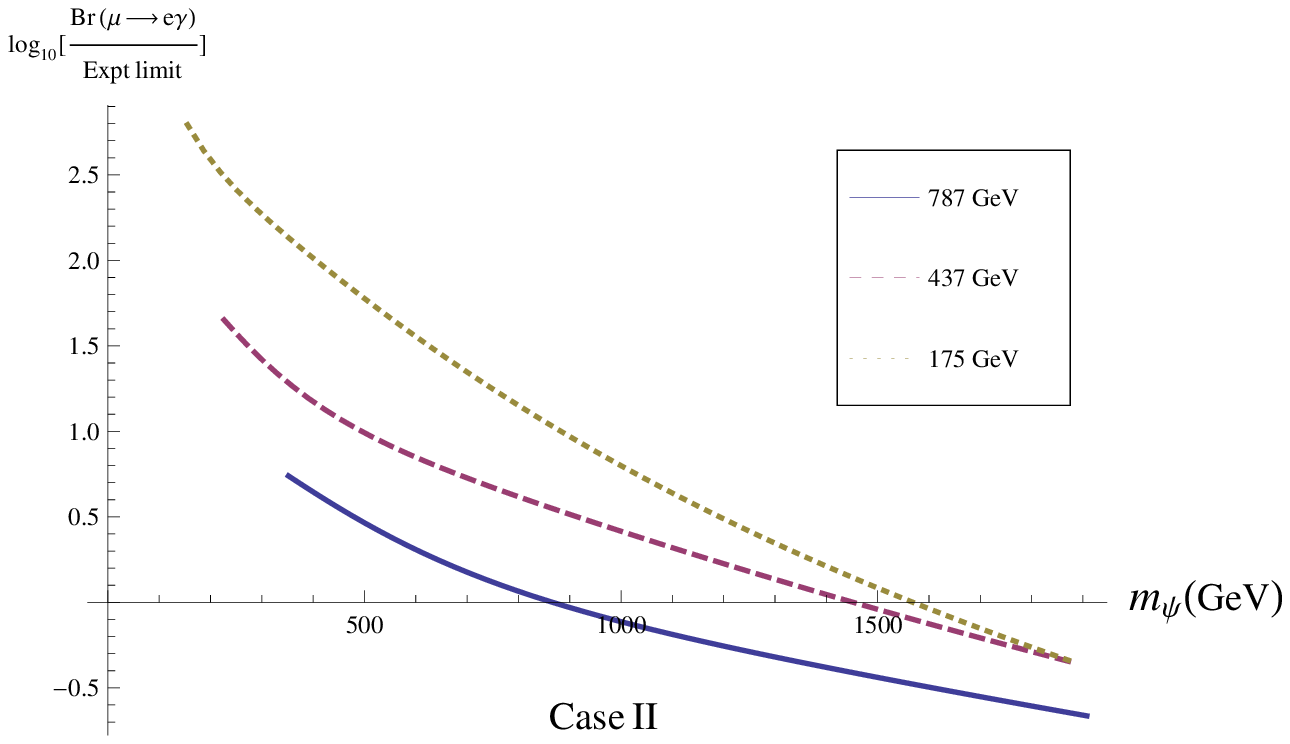}}
{\includegraphics[width=10 cm]{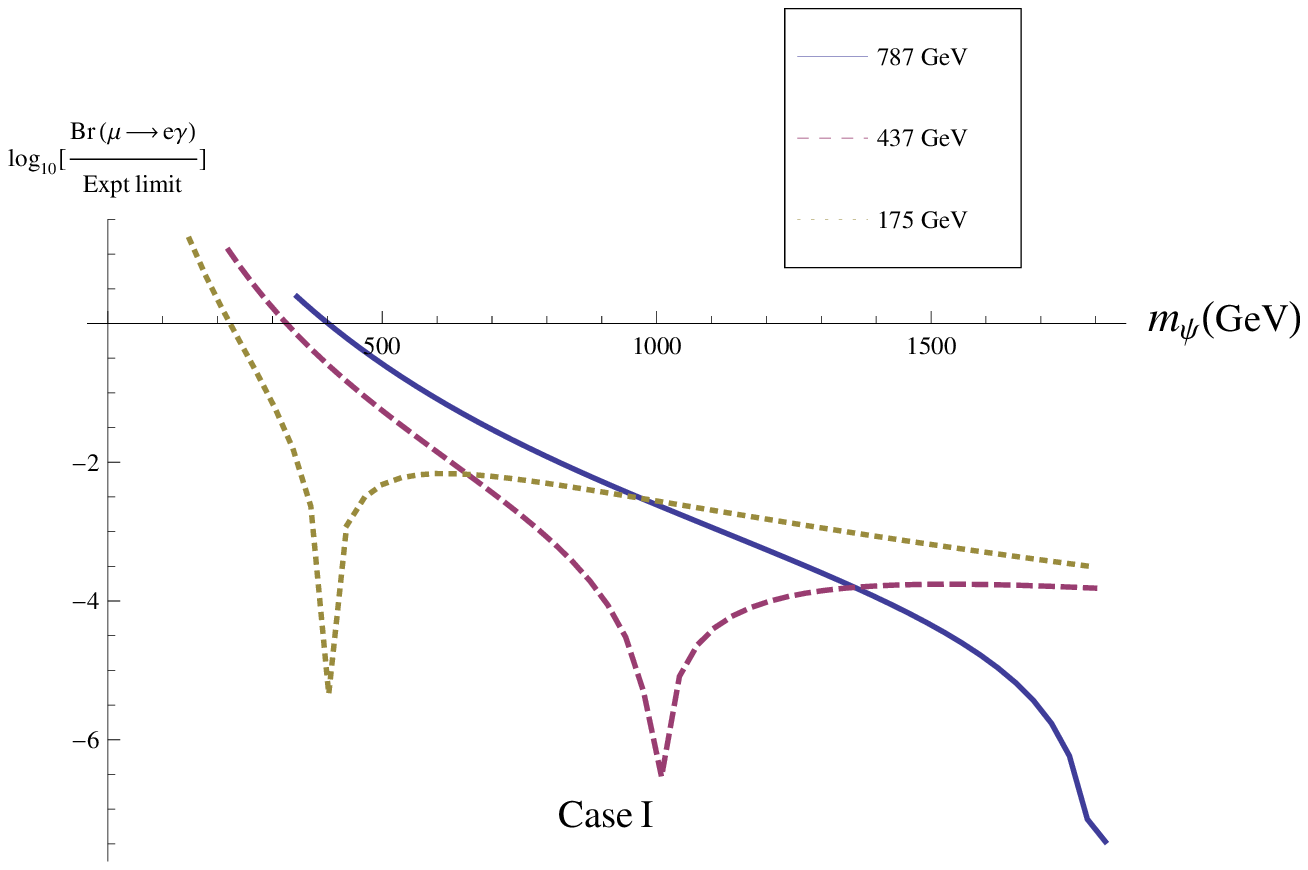}} \caption{ The above graphs
show the plot of Log of Br($\mu\rightarrow e\gamma$) divided by
experimental bound ($1.2\times 10^{-11}$) versus $m_{\psi}$ for two
cases I and II with $M_{1/2}$=787 GeV, 437 GeV and 175 GeV.}
\label{p3mu}
\end{figure}

The Dirac neutrino Yukawa couplings $(Y_N)_{ij}$ induce flavor
violating off-diagonal elements in the left-handed slepton mass
matrix through the RG running from $M_{\rm{GUT}}$ to the
right-handed neutrino mass scales. The RGEs for MSSM with
right-handed neutrinos are given in Ref.\cite{RGEs}. The
right-handed neutrinos $M_{R_i}$ are determined in the $SO(10)\times
A_4$ model. In this case, the induced mass insertion parameters for
left-handed sleptons are given by \cite{BPR},
\begin{eqnarray}
(\delta^l_{LL})_{ij}^{RHN}=-\frac{3m^2_{\psi}+\tilde{a}^2}{8m^2_{\psi}\pi^2}\sum_{k=1}^{3}
(Y_N)_{ik}(Y^*_N)_{jk}ln\frac{M_{GUT}}{M_{R_k}},
\end{eqnarray}
where the matrix $Y_N$ is written in the mass eigenstates of charged
leptons and right-handed neutrinos. The total LL contribution for
the charged leptons is given by
\begin{eqnarray}
(\delta^l_{LL})_{ij}^{Tot}=(\delta^l_{LL})_{ij}^{RHN}+(\delta^l_{LL})_{ij}.
\end{eqnarray}

\section{Results}

In this section, we investigate the flavor violating processes by
calculating the mass insertion parameters $\delta_{LL}$,
$\delta_{RR}$, and $\delta_{LR,RL}$, then we compare them with their
experimental bounds. These bounds in the quark and lepton sectors
were obtained by comparing the hadronic and leptonic flavor changing
processes to their experimental values/limits \cite{3p3,4p3}.
Eq.(\ref{p3LL1}), Eq.(\ref{p3LL2}) and Eq.(\ref{p3LL3}) are used to
calculate $\delta_{LL,RR}$ and Eq.(\ref{p3LR}) is used to calculate
$\delta_{LR,RL}$ for both charged leptons and down quarks. The
result of mass insertion calculations and their experimental bounds
are presented in Table \ref{p3inser3}. In this table, we have
defined $\sigma=\frac{\tilde m^2_{{\chi}_{2}}-\tilde
m^2_{{\psi}_{i}}}{\tilde m^2_{{\psi}_{i}}}$ and $\tilde k= \tilde
zm_{b,\tau}$
\begin{table}[t]
\center
\begin{tabular}{|c|c|c|c|} \hline
Mass Insertion ($\delta$) & Model Predictions & Exp. upper Bounds\\
\hline
$(\delta^l_{12})_{LL}$ &     0.062$~\sigma$+$(\delta^l_{12})_{LL}^{RHN}$ &  6 $\times$ $10^{-4}$    \\
$(\delta_{12}^l)_{RR}$  & 6.1 $\times$ $10^{-4}$$~\sigma$&   0.09        \\
$(\delta^l_{12})_{RL,LR}$ &  (0.084,~0.0096)$~\tilde k$      &   $10^{-5}$          \\
$(\delta^l_{13})_{LL}$    &    0.022$~\sigma$+$(\delta^l_{13})_{LL}^{RHN}$ &  0.15        \\
$(\delta^l_{13})_{RR}$    &    0.028$~\sigma$  &   -         \\
$(\delta^l_{13})_{RL,LR}$& (0.0335,~0.076)$~\tilde k$ &  0.04           \\
$(\delta^l_{23})_{LL}$    &   0.27$~\sigma$+$(\delta^l_{13})_{LL}^{RHN}$   &    0.12         \\
$(\delta^l_{23})_{RR}$    &    0.034$~\sigma$  &     -        \\
$(\delta^l_{23})_{RL,LR}$&   (0.055,~0.899)$~\tilde k$  &      0.03
\\ \hline \hline
$(\delta^d_{12})_{LL}$ &1.9 $\times$ $10^{-4}$$~\sigma$ &  0.014                \\
$(\delta_{12}^d)_{RR}$  &  0.15$~\sigma$     &     0.009                    \\
$(\delta^d_{12})_{LR,RL}$ &(0.029,~0.035)$~\tilde k$  &  $9 \times 10^{-5}$    \\
$(\delta^d_{13})_{LL}$    &  0.014$~\sigma$    &   0.09          \\
$(\delta^d_{13})_{RR}$    &  0.061$~\sigma$    &   0.07          \\
$(\delta^d_{13})_{LR,RL}$&(0.173,~0.016)$~\tilde k$  & $1.7 \times 10^{-2}$ \\
$(\delta^d_{23})_{LL}$    &   0.054$~\sigma$  &    0.16         \\
$(\delta^d_{23})_{RR}$    & 0.29$~\sigma$      &    0.22         \\
$(\delta^d_{23})_{LR,RL}$& (0.875,~0.064)$~\tilde k$ &
(0.006,~0.0045)     \\\hline
\end{tabular}
\caption{The mass insertion parameters predicted by $SO(10)\times
A_4$ model and their experimental upper bounds obtained from
\cite{4p3}.}\label{p3inser3}
\end{table}

The stringent bounds on leptonic $\delta_{12}$, $\delta_{13}$, and
$\delta_{23}$ in Table \ref{p3inser3} come only from the decay rates
$l_i\rightarrow l_j\gamma$. The experimental bounds on the mass
insertion parameters listed in column 3 were obtained by making a
scan of $m_0$ and $M_{1/2}$ over the ranges $m_0<380~GeV$ and
$M_{1/2}<160~GeV$, where $m_0$ and $M_{1/2}$ are the scalar
universal mass and the gaugino mass respectively \cite{4p3}.

Glancing at Table \ref{p3inser3}, we note that the stringent
constraint on leptonic flavor violation arises from $\delta^l_{12}$
which corresponds to the decay rate of $\mu\rightarrow e\gamma$. On
the other hand, there is a weaker constraint that arises from
$\delta^d_{12}$ on the quark sector. One can do an arrangement such
that $\tilde a-\tilde b_3=200$ GeV and $\tilde m_f= 800$ GeV
(equivalent to $\tilde k=2.6\times 10^{-4}$) so that all the
chirality flipping mass insertions will be within their experimental
bounds. This arrangement is possible if the trilinear soft terms
vanish at the scale $M^*$.

Since the stringent constraint comes from the $\mu\rightarrow
e\gamma$ process, let us discuss the branching ratio of this process
in more details. In general, the branching ratio of $\l_i\rightarrow
\l_j\gamma$ is given by
\begin{eqnarray}
\frac{BR(l_{i}\rightarrow  l_{j}\gamma)}{BR(l_{i}\rightarrow
l_{j}\nu_i\bar{\nu_j})} =
\frac{48\pi^{3}\alpha}{G_{F}^{2}}(|A_L^{ij}|^2+|A_R^{ij}|^2).
\end{eqnarray}
We have used the general expressions for the amplitudes
$A_{L,R}^{ij}$ given by Ref.\cite{Paradisi:2005fk} where the
contributions from both chargino and the neutralino loops are
included. These expressions are written in terms of mass insertion
parameters.

The correct suppression of the decay  rate $\Gamma(\mu\rightarrow
e\gamma)$  requires a high degree of degeneracy of the soft mass
terms of MSSM fields and the exotic fields. For example,
$\sigma\approx 0.01$, as can be seen from Table \ref{p3inser3}. In
order to obtain high degree of degeneracy, let us assume that the
SSB terms which are generated at the messenger scale $M^*$ satisfy
the universality boundary conditions at the scale $M^*$ given by
\begin{eqnarray}
\tilde{m}^2_{\psi_i}&=&\tilde{m}^2_{\chi_i}=\tilde{m}^2_{\overline{\chi}_i}=\tilde{m}^2_{10_H}=\tilde{m}^2_{1_H}=\tilde{m}^2_{1'_H}=m_0 ,\nonumber\\
M_{\lambda}&=&M_{0}, \nonumber\\
\tilde{a}&=&\tilde{b}_1=\tilde{b}_2=0, \label{p3bc}
\end{eqnarray}
where $M_{\lambda}$ is the gaugino mass of $SO(10)$ gauge group.
Solving the RGE listed in Appendix C with the boundary conditions
given by Eq.(\ref{p3bc}) determines the value of $\sigma$. In Table
\ref{p3tabr} we give the branching ratio of the process
$\mu\rightarrow e\gamma$ predicted by the $SO(10)\times A_4$ model
for different choices of the input parameters $a$, $b_1$, $b_2$,
$\tilde{m}_{\psi}$ and $M_{1/2}$ at the GUT scale. The experimental
searches have put the upper limit on the branching ratio of
$\mu\rightarrow e\gamma$ as $\rm{Br}(\mu\rightarrow e\gamma)\leq
1.2\times 10^{-11}$ \cite{Br}. Note that $\tilde{m}_{\psi}$ and
$M_{1/2}$ originate respectively from $m_0$ and $M_0$ through RGEs.
In this Table we consider $\ln\frac{M^*}{M_{GUT}}=1$ and
$\ln\frac{M^*}{M_{GUT}}=4.6$ that correspond respectively to
$M^*\approx 3M_{\rm{GUT}}$ and $M^*\approx M_{\rm{Planck}}$.

\begin{table}[t]
\center
\begin{tabular}{|c|c|c|c|c|} \hline
   &I& II& III & IV \\\hline
$a$  &     1.14       &  1.07      &  1.14 &         0.62
\\\hline
 $b_1$  &1.9 &1.5 &     1.24  & 1.24                                           \\\hline
 $b_2$  &1.9 & 1.5&       1.24 & 1.24                                            \\\hline
 $\tilde{m}_{\psi_i}$  & 542&886 &  2932 &675           \\\hline
 $M_{1/2}$ &350 &787 & 1924 & 350                                      \\\hline
 BR($\mu\rightarrow e\gamma$) & $1.4\times10^{-13}$ & $1.16\times10^{-11}$ &
 $1.2\times10^{-11}$& $2.2\times10^{-12}$\\ \hline
\end{tabular}
\caption{Branching ratio of $\mu\rightarrow e\gamma$ for different
choices of input parameters at the GUT scale. Cases I and II
correspond to $\ln\frac{M^*}{M_{GUT}}=1$ and cases III and IV
correspond to $\ln\frac{M^*}{M_{GUT}}=4.6$. $\tilde{m}_{\psi_i}$ and
$M_{1/2}$ are given in GeV} \label{p3tabr}
\end{table}
Let us analyze the four cases in the Table \ref{p3tabr}. In the
cases (I, II and III), the chosen values of the parameters $a$ are
consistent with the top Yukawa coupling of order $0.5$ at the GUT
scale and with the fitting for fermion masses and mixing. On the
other hand, the choice of $a=0.68$ in Case IV is not consistent with
the fit. Although the medium slepton masses of order $550$ GeV are
obtained in Case I, the choice $b_1=b_2=1.9$ corresponds to
non-perturbative Yukawa couplings at the scale  $M^*$ (i.e.
$b_1=b_2=4$ at $M^*$). In this case, the solutions of the 1-loop
RGEs are not trusted since the Yukawa couplings $b_1$ and $b_2$ go
non-perturbative above the GUT scale. Also, it is important to point
out that the flavor violation constraint on $\mu\rightarrow e\gamma$
in Case III requires heavy slepton masses ($\geq 3$ TeV) while it
requires slepton masses of order $\sim 900$ GeV in Case II. In other
words, Case II is preferred in our model in the sense that the decay
rate of $\mu\rightarrow e\gamma$ is close to the experimental limit
with a reasonable supersymmetric mass spectrum, so it might be
tested in the ongoing MEG experiment\cite{MEG}. Besides, the Yukawa
couplings remain perturbative at the messenger scale $M^*$. Figure
\ref{p3mu} shows the allowed values of $m_{\psi}$ that correspond to
the graphs below the $x$-axis for the cases I and II.

\section{Conclusion}

In this paper, we investigated flavor violating processes that arise
below and above the GUT scale in the $SO(10)\times A_4$ model. Above
the GUT scale, we study how flavor violation gets linked with the
fitting of fermion masses and mixing through the factors $T_1$,
$T_2$, and $T_3$. The requirement of top Yukawa coupling being
$\sim$ 0.5 necessitates some of these factors to be large.
Consequently, this corresponds to an order one mixing of the light
fields with the exotic heavy fields. In this case, flavor violation
is reintroduced at the GUT scale where $A_4$ symmetry is broken. The
stringent constraint on $\mu\rightarrow e\gamma$ decay rate requires
a high degree of degeneracy of the soft quadratic masses of the
exotic heavy fields and the light fields. Therefore, all the
quadratic soft masses are assumed to be universal at the scale
$M^*\sim 3M_{GUT}$. Flavor violation is also induced below the GUT
scale in the presence of right handed neutrinos through the RG
running from $M_{\rm{GUT}}$ to the right handed neutrino mass
scales. This FV source is predicted by $SO(10)\times A_4$ model
because the Dirac neutrino Yukawa couplings are determined from the
fermion masses and mixings fitting. Combining all sources of FV, we
found that Case I and Case II presented in Table \ref{p3tabr} are
consistent with fermion masses and mixing fitting and with
$\mu\rightarrow e\gamma$ decay rate, which is however predicted to
be close to the current experimental bound. Thus the ongoing MEG
experiment can confirm or rule out our model. Case I that
corresponds to slepton mass of order 1 TeV is also consistent with
the Yukawa couplings (i.e., $b_1$ and $b_2$) being perturbative at
the scale $M^*$. On the other hand, these Yukawa couplings do not
remain perturbative at $M^*$ in Case II that corresponds to slepton
masses of order $550$ GeV.

\section*{Acknowledgements}
I would like to thank Professor K. S. Babu for reading the
manuscript and valuable suggestions. This work is supported by US
Department of Energy, Grant Number DE-FG02-ER46140.
\section{Appendices}
\appendix
\section{Derivation of the Light Fermion Mass Matrix}
In order to block-diagonalize the mass matrix of $W_{spinor}$, we
define the new orthogonal basis as $Y=UX$, where Y(X) is the column
matrix that contains the new(old) eigenstates and $U$ is the
$5\times5$ orthogonal matrix (i.e $U^TU=UU^T=I$). These matrices are
given by:
\begin{eqnarray}
 \footnotesize\left(\begin{array}{c} L_2 \\ L_3 \\ H_1 \\ H_2 \\ H_3
\end{array}\right) = \footnotesize\left(
\begin{array}{ccccc}
-N_1 & 0 & 0&N_1s_{\theta}T_2&0 \\
\frac{fN_1c_{\theta}s_{\theta}T^2_2}{r_F}&-\frac{f}{N_1r_F}  & \frac{fT_1}{N_1r_F}&\frac{fN_1c_{\theta}T_2}{r_F}&-\frac{fQ_FT_1T_3}{N_1r_F} \\
0 & 0 & G_FQ_FT_3&  0 & G_F\\
N_2s_{\theta}T_2 & N_2c_{\theta}T_2 & 0&  N_2 & 0\\
\frac{fN_2c_{\theta}s_{\theta}T^2_2T_1}{G_Fr_F}&-\frac{fN_2T_1}{N^2_1G_Fr_F}
&-\frac{fG_F}{N_2r_F}& \frac{fN_2c_{\theta}T_2T_1}{G_Fr_F} &
\frac{fQ_FG_FT_3}{N_2r_F}
\end{array} \right) \footnotesize\left(\begin{array}{c} \psi_2 \\ \psi_3 \\ \chi_1 \\ \chi_2 \\ \chi_3
\end{array}\right), \label{p3bc1}
\end{eqnarray}
where $N_1=1/ \sqrt{1+T^2_2s^2_{\theta}}$, $N_2=1/ \sqrt{1+T^2_2}$,
$G_F=1/ \sqrt{1+T^2_3Q^2_{F}}$,
$f=(1+T^2_2+T^2_1(1+s^2_{\theta}T^2_2))^{-1/2}$ , and
$r_F=\sqrt{(1+Q^2_FT^2_3T^2_1(1+s^2_{\theta}T^2_2)f^2)}$. The
parameters appearing in the above matrix are assumed to be real.
Define $e_i$, $E_{i}$, $\overline{E^c}_{i}$, $g_i$, $g'_i$, $g''_i$,
and $g'''_i$ to be the charge ($-1$) leptons in the $\psi_i$,
$\chi_{i}$, $\overline{\chi}_{i}$, $\phi_i$, $\phi'_i$, $\phi''_i$,
and $\phi'''_i$, respectively; and define $e^c_i$, $E^c_{i}$,
$\overline{E}_{i}$, $g^c_i$, $g'^c_i$, $g''^c_i$, and $g'''^c_i$ to
be the charge ($+1$) antileptons in the same representations. By
writing the old eigenstates appearing in the superpotential
($W_{spin}+W_{vect}$) of Eqs.(\ref{p32}) and (\ref{p33}) in terms of
the new ones, and restricting attention to the electron-type
leptons, one gets a $21\times21$ mass matrix:
\begin{eqnarray}
W_{mass}=\left(\begin{array}{ccccccc} e^c_i& E^c_{\alpha}&
\overline{E}_{\alpha}& g^c_i& g'^c_i& g''^c_i&  g'''^c_i
\end{array}\right) \left(
\begin{array}{cc}
m_0 & m \\
M'  & M\end{array} \right)\left(\begin{array}{c} e_i\\ E_{\alpha}\\
\overline{E^c}_{\alpha}\\ g_i\\ g'_i\\ g''_i\\ g'''_i
\end{array}\right) ,
\end{eqnarray}
where,
\begin{eqnarray}
m_0= \left(\begin{array}{ccc}
 0 & 0 & 0 \\
 0 & 0 & -\frac{a f v_d Q_e s_{\theta } T_1 T_2 T_3}{r_e} \\
 0 & -\frac{a f v_d Q_{e^c} s_{\theta } T_1 T_2 T_3}{r_{e^c}} & -\frac{a f^2 v_d \left(Q_e+Q_{e^c}\right) c_{\theta } T_1 T_2 T_3}{r_e
r_{e^c}}
\end{array}\right)\nonumber.
\end{eqnarray}
The matrices $M'$, $m$ and $M$ can be written in the compact form as

\begin{eqnarray}
M'&=&\left(
\begin{array}{c}
 M'_{11}\\
 0
\end{array}
\right),\\
m^T&=&\left(\begin{array}{c}
 m_{11} \\
 0
\end{array}
\right),\\
 M&=&\left(\begin{array}{ccc}
 M_{11}&M_{12}&M_{13}
\end{array}\right),
\end{eqnarray}
where
\begin{eqnarray}
 M'_{11}&=&\left(
\begin{array}{ccc}
 0 & a N_1v_d G_{e^c} s_{\theta } T_2 & \frac{a f N_1 v_d G_{e^c} c_{\theta } T_2}{r_e} \\
 0 & 0 & -\frac{a f N_2 v_d Q_e T_1 T_3}{N_1 r_e} \\
 0 & \frac{a f N_1 v_d G_{e^c} Q_{e^c} s_{\theta } T_2 T_3}{N_2 r_{e^c}} & \frac{a f^2 v_d c_{\theta }\left(N_1^2 G_{e^c}^2
Q_{e^c}-N_2^2 Q_e T_1^2\right) T_2 T_3}{N_1 N_2 G_{e^c} r_e r_{e^c}} \\
 0 & 0 & 0 \\
 0 & 0 & 0 \\
 0 & 0 & 0 \\
 b_3 v_1 & 0 & 0 \\
 0 & -b_3 N_1 v_1 & \frac{b_3 f N_1 v_1 s_{\theta } c_{\theta } T_2^2}{r_e} \\
 0 & 0 & -\frac{b_3 f v_1}{N_1 r_e}
\end{array}
\right),\nonumber\\
m_{11}&=&\left(\begin{array}{ccc}
 0 & a N_1 v_d G_e s_{\theta } T_2 & \frac{a f N_1 v_d c_{\theta } G_e T_2}{r_{e^c}} \\
 0 & 0 & -\frac{a f N_2 v_d Q_{e^c} T_1 T_3}{N_1 r_{e^c}} \\
 0 & \frac{a f N_1 v_d G_e Q_e s_{\theta } T_2 T_3}{N_2 r_e} & \frac{a f^2 v_d c_{\theta } \left(N_1^2 G_e^2 Q_e-N_2^2
Q_{e^c} T_1^2\right) T_2 T_3}{N_1 N_2 G_e r_e r_{e^c}} \\
 0 & 0 & 0 \\
 0 & 0 & 0 \\
 0 & 0 & 0 \\
 b_3 v_5 & 0 & 0 \\
 0 & -b_3 N_1 v_5 & \frac{b_3 f N_1 v_5 c_{\theta } s_{\theta } T_2^2}{r_{e^c}} \\
 0 & 0 & -\frac{b_3 f v_5}{N_1 r_{e^c}}
\end{array}
\right)\nonumber,
\end{eqnarray}

\begin{eqnarray*}
M_{11}=\footnotesize\left(\begin{array}{cccc}
0                                                               & a N_2 v_d G_{e^c}                             & \frac{a f N_2 v_d G_{e^c} c_{\theta }T_1 T_2}{G_e r_e}                                                    & M_1 G_{e^c} Q_{e^c} T_3                                                           \\
a N_2 v_d G_e                                                   & 0                                             & \frac{a f v_d G_e Q_e T_3}{r_e}                                                                           & M_1 N_2 c_{\theta } T_1 T_2                                                       \\
\frac{a f N_2 v_d G_e c_{\theta } T_1 T_2}{G_{e^c} r_{e^c}}     & \frac{a f v_d G_{e^c} Q_{e^c} T_3}{r_{e^c}}   & \frac{a f^2 v_d\left(G_e^2 Q_e+G_{e^c}^2 Q_{e^c}\right) c_{\theta } T_1 T_2T_3}{G_e G_{e^c} r_e r_{e^c}}  & -\frac{f M_1 \left(N_1^2G_{e^c}^2+N_2^2 T_1^2\right)}{N_1^2 N_2 G_{e^c}r_{e^c}}   \\
M_1 G_e Q_e T_3                                                 & M_1 N_2 c_{\theta } T_1 T_2                   & -\frac{f M_1 \left(N_1^2 G_e^2+N_2^2 T_1^2\right)}{N_1^2N_2 G_e r_e}                                      & 0                                                                                 \\
0                                                               & \frac{M_2 }{N_2}                              & 0                                                                                                         & 0                                                                                 \\
\frac{M_3 }{ G_e}                                               & 0                                             & 0                                                                                                         & 0                                                                                 \\
0                                                               & 0                                             & 0                                                                                                         & 0                                                                                 \\
0                                                               & b_3 N_2 v_1 s_{\theta } T_2                   & \frac{b_3 f N_2 v_1 s_{\theta } c_{\theta }T_1 T_2^2}{G_e r_e}                                            & 0                                                                                 \\
0                                                               & b_3 N_2 v_1 c_{\theta } T_2                   & -\frac{b_3 f N_2 v_1 T_1}{N_1^2 G_e r_e}                                                                  & 0                                                                                 \\
0                                                               & 0                                             & 0                                                                                                         & 0                                                                                 \\
0                                                               & 0                                             & 0                                                                                                         & 0                                                                                 \\
0                                                               & 0                                             & 0                                                                                                         & 0                                                                                 \\
0                                                               & 0                                             & 0                                                                                                         & 0                                                                                 \\
0                                                               & 0                                             & 0                                                                                                         & 0                                                                                 \\
0                                                               & 0                                             & 0                                                                                                         & 0                                                                                 \\
0                                                               & 0                                             & 0                                                                                                         & 0                                                                                 \\
0                                                               & 0                                             & 0                                                                                                         & 0                                                                                 \\
0                                                               & 0
& 0 & 0
\end{array}\right),\nonumber
\end{eqnarray*}
\begin{eqnarray}
M_{12}=\footnotesize\left(\begin{array}{ccccccccccc}
0 & \frac{M_3}{ G_{e^c}} & 0 & 0 & 0 & 0 & 0 \\
\frac{M_2 }{N_2} &0 & 0 & b_3 N_2 v_5 s_{\theta } T_2 & b_3 N_2 v_5 c_{\theta } T_2 & 0 & 0 \\
0 & 0 & 0 & \frac{b_3 f N_2 v_5s_{\theta } c_{\theta } T_1 T_2^2}{G_{e^c}r_{e^c}} & -\frac{b_3 f N_2 v_5 T_1}{N_1^2 G_{e^c} r_{e^c}} & 0 & 0 \\
0 & 0 & 0 & 0 & 0 & 0 & 0  \\
0 & 0 & 0 & 0 & 0 & 0 & 0   \\
0 & 0 & 0 & 0 & 0 & 0 & 0   \\
0 & 0 & 0 & 0 & 0 & M_{10} & 0 \\
0 & 0& 0 & 0 & 0 & 0 & M_{10}  \\
0 & 0 & 0 & 0 & 0 & 0& 0   \\
0 & 0 & M_{10} & 0 & 0 & 0 & h \epsilon_3   \\
0 & 0 & 0 & M_{10} & 0 & h \epsilon_3 & 0   \\
0 & 0 & 0 & 0 & M_{10} & h \epsilon_2 & h\epsilon_1 \\
0 & 0 & 0 & 0 & 0 & 0 & A_2\gamma_3  \\
0 & 0 & 0 & 0 & 0 & A_1\gamma_3 & 0  \\
0 & 0 & 0 & 0 & 0 & A_2\gamma_2 & A_1\gamma_1  \\
0 & 0 & 0 & 0 & 0 & k_2 \Omega  Q_e & 0  \\
0 & 0 & 0 & 0 & 0 & 0 & k_2 \Omega  Q_e    \\
0 & 0 & 0 & 0 & 0 & 0 & 0
\end{array}\right),\nonumber
\end{eqnarray}
 and
\begin{eqnarray}
M_{13}=\footnotesize\left(\begin{array}{ccccccc}
0 & 0 & 0 & 0 & 0 & 0 & 0 \\
0 & 0 & 0 & 0 & 0 & 0 & 0 \\
0 & 0 & 0 & 0 & 0 & 0 & 0 \\
0 & 0 & 0 & 0 & 0 & 0 & 0 \\
0 & 0 & 0 & 0 & 0 & 0 & 0 \\
0 & 0 & 0 & 0 & 0 & 0 & 0 \\
0 & 0 & 0 & 0 & 0 & 0 & 0 \\
0 & 0 & 0 & 0 & 0 & 0 & 0 \\
M_{10}& 0 & 0 & 0 & 0 & 0 & 0 \\
h\epsilon_2 & 0 & A_1\gamma_3&A_2\gamma_2 & -k_2 \Omega  Q_e & 0 & 0 \\
h \epsilon_1 & A_2\gamma_3 & 0&A_1\gamma_1 & 0 & -k_2 \Omega  Q_e & 0 \\
0 & A_1\gamma_2 & A_2\gamma_1 & 0 & 0 & -k_2 \Omega  Q_e \\
A_1\gamma_2 & 0 & 0& 0 & m & 0 & 0 \\
A_2\gamma_1 & 0 & 0& 0 & 0 & m & 0 \\
0 & 0 & 0 & 0 & 0 & 0 & m \\
0 & m & 0 & 0 & 0 & 0 & 0 \\
0 & 0 & m & 0 & 0 & 0 & 0 \\
k_2 \Omega  Q_e &0 & 0 & m & 0 & 0 & 0
\end{array}\right).\nonumber
\end{eqnarray}
Here $v_1=\langle1(16_H)\rangle$,
$v_5=\langle\overline{5}(16_H)\rangle$,
$v_d=\langle\overline{5}(10_H)\rangle$, $s_{\theta}\equiv\sin\theta$
and $c_{\theta}\equiv\cos\theta$. The above $21\times 21$ mass
matrix may be block-diagonalized as follows \cite{babu-barr}:
\begin{eqnarray}
U_R\footnotesize \left(
\begin{array}{cc}
m_0 & m \\
M'  & M\end{array} \right)U^{\dag}_L=\left(
\begin{array}{cc}
(m_0 -mM^{-1}M')(1+y^{\dag}y)^{-1/2}&0\\
0  & (MM^{\dag}+M'M'^{\dag})\end{array} \right)\label{p310},
\end{eqnarray}
where
\begin{eqnarray}
U_R =\footnotesize\left(
\begin{array}{cc}
I&(m_0M'^{\dag} +mM^{\dag})(MM^{\dag}+M'M'^{\dag})^{-1}\\
(MM^{\dag}+M'M'^{\dag})^{-1}(m^{\dag}_0M' +m^{\dag}M)  & I
\end{array} \right)\label{p3t1},
\end{eqnarray}
and
\begin{eqnarray}
U_L =\footnotesize\left(
\begin{array}{cc}
(1+y^{\dag}y)^{-1/2}&0\\
0& (MM^{\dag}+M'M'^{\dag})^{-1/2}\end{array} \right)\left(
\begin{array}{cc}
I&-y^{\dag}\\
M'& M\end{array} \right)\label{p3t2}.
\end{eqnarray}
Here $y=M^{-1}M'$. Terms of order $(M_{Weak}/M_{GUT})^2$ have been
dropped. Then the $3\times3$ light fermion mass matrix of charged
leptons in Eq.(\ref{p31}) is obtained by applying the relation in
the left upper block of the matrix in the Eq.(\ref{p310}), where the
factor $(1+y^{\dag}y)^{-1/2}$ is close to identity for small mixing
between the $\psi_i$ and the 10-plet vectors. Similarly, one can
obtain the down-type quark mass matrix. The parameters appearing in
Eq.(\ref{p31}) are defined as follows:
\begin{eqnarray}
\zeta&=&c_{13}+\delta_2\frac{3+2\alpha}{5},\\
\beta&=&c_{23}-\delta_1\frac{3+2\alpha}{5},\\
s&=&\frac{5s_{\theta }}{f(2+3\alpha )c_{\theta}},\\
c_{12}&=&\frac{b_3^2 h N_1 v_1 v_5 \epsilon_3}{a f^2 v_d c_{\theta }
M_{10}^2(Q_e+Q_{e^c}) T_1 T_2 T_3},\\
\delta_3&=&\frac{(A_1-A_2) b_3^2 k_2 N_1 v_1 v_5 \gamma_3 \Omega }{a
f^2 m v_d c_{\theta } M_{10}^2 \left(Q_e+Q_{e^c}\right)
T_1 T_2 T_3},\label{p3d3}\\
c_{13}&=&\frac{b_3^2 h v_1 v_5\left(\epsilon_2-N_1^2 \epsilon_3
c_{\theta } s_{\theta }
T_2^2\right)}{a f N_1 v_d c_{\theta } M_{10}^2 \left(Q_e+Q_{e^c}\right) T_1 T_2 T_3},\\
\delta_2&=&\frac{(A_1-A_2) b_3^2 k_2 v_1 v_5 \Omega \rm{
}\left(\gamma_2+N_1^2
\gamma_3 c_{\theta } s_{\theta } T_2^2\right)}{a f m N_1 v_d c_{\theta } M_{10}^2 \left(Q_e+Q_{e^c}\right) T_1 T_2 T_3},\label{p3d2}\\
c_{23}&=&\frac{-b_3^2 h v_1 v_5\epsilon_1}{a f v_d c_{\theta } M_{10}^2(Q_e+Q_{e^c}) T_1 T_2 T_3},\\
\delta_1&=&\frac{(-A_1+A_2) b_3^2 k_2 v_1 v_5 \gamma_1 \Omega  }{a f
m v_d c_{\theta } M_{10}^2 \left(Q_e+Q_{e^c}\right) T_1 T_2
T_3}\label{p3d1}.
\end{eqnarray}
The above parameters are written in terms of the Yukawa couplings
and the VEVs of the Higgs fields appearing in the superpotentials
$W_{spin}$ and $W_{vect.}$ in Eqs.(\ref{p32}) and (\ref{p33}). The
parameters $\gamma_1$, $\gamma_2$ and $\gamma_3$ appearing in the
above Eqs.(\ref{p3d3}), (\ref{p3d2}) and (\ref{p3d1}) are the VEV
components of the Higgs singlet $1''_H$.

\section{Light Neutrino Mass Matrix}
The neutrino mass matrix can be obtained from the superpotentials
given by Eqs.(\ref{p32}) and (\ref{p34}). For simplicity, the
contribution from the superpotential $W_{vect.}$ in Eq.(\ref{p33})
is ignored by assuming the coupling of the ordinary spinor fields
$16_i$ with the vector multiplets is small. Define the right- and
left-handed neutrinos, denoted respectively by ($\nu^c_i$ and
$\nu_i$), residing in $\psi_i$. Similarly, $\nu^c_{\chi_i}$ and
$\nu_{\chi_ i}$($\overline{\nu^c}_{\overline{\chi}_i}$ and
$\overline{\nu}_{\overline{\chi}_i}$) reside in
$\chi_{i}$($\overline{\chi}_{i}$) where $i$ runs from 1 to 3.
Including the six singlets denoted by $Z_i$ and $Z_i^c$, one can
construct $24\times24$ mass matrix written in the following compact
form
\begin{eqnarray}
W_{mass}=N^{T}\left(\begin{array}{cc}
0&M_D\\
M_D^T& M_R\end{array} \right)N,
\end{eqnarray}
where
\begin{eqnarray}
N^{T}= \left(\begin{array}{cccccccc} \nu_i & \nu_{\chi_ i}&
\overline{\nu^c}_{\overline{\chi}_i}& \nu^c_i&\nu^c_{\chi_i}&
\overline{\nu}_{\overline{\chi}_i}& Z_i & Z^c_i\end{array}\right),
\end{eqnarray}
and
\begin{eqnarray}
M_D^T=\left(
\begin{array}{c}
0\\
C\\
0
\end{array}
\right),
\end{eqnarray}
where
\begin{eqnarray}
C=\left(
\begin{array}{ccc}
 0 & 0 & -\frac{a f v_u Q_{\nu} s_{\theta } T_1 T_2 T_3}{r_{\nu}} \\
 0 & -\frac{a f v_u Q_{{\nu}^c} s_{\theta } T_1 T_2 T_3}{r_{{\nu}^c}} & -\frac{a f^2 v_u c_{\theta } \left(Q_{\nu}+Q_{{\nu}^c}\right) T_1 T_2 T_3}{r_{\nu} r_{{\nu}^c}} \\
 0 & a N_1 v_u G_{{\nu}^c} s_{\theta } T_2 & \frac{a f N_1 v_u
  c_{\theta } G_{{\nu}^c} T_2}{r_{\nu}} \\
 0 & 0 & -\frac{a f N_2 v_u
  Q_{\nu} T_1 T_3}{N_1 r_{\nu}} \\
 0 & \frac{a f N_1 v_u
  G_{{\nu}^c} Q_{{\nu}^c} s_{\theta } T_2 T_3}{N_2 r_{{\nu}^c}} & \frac{a f^2 v_u
   c_{\theta } \left(N_1^2 G_{\nu^c}^2 Q_{\nu^c}-N_2
 ^2 Q_{\nu} T_1^2\right) T_2 T_3}{N_1 N_2
  G_{\nu^c} r_{\nu} r_{\nu^c}}
\end{array}
\right).\nonumber
\end{eqnarray}
Here $v_u=\langle 5(10_H)\rangle$. The matrix $M_R$ can be written in the compact form
\begin{eqnarray}
M_R=\left(
\begin{array}{cccc}
M_{R11}&M_{R12}&M_{R13}&M_{R14}
\end{array}
\right),
\end{eqnarray}
where the matrices  $M_{R11}$, $M_{R12}$, $M_{R13}$, and  $M_{R14}$
are given respectively by

\begin{eqnarray}
\footnotesize \left(
\begin{array}{cccc}
 0 & 0 & 0 & 0 \\
 0 & 0 & 0 & 0 \\
 0 & 0 & 0 & 0 \\
 0 & 0 & 0 & 0 \\
 0 & 0 & 0 & 0 \\
 0 & 0 & 0 & 0 \\
 0 & 0 & 0 & 0 \\
 a N_1 v_u G_{\nu} s_{\theta } T_2 & 0 & \frac{a f N_1 v_u G_{\nu} Q_{\nu} s_{\theta } T_2 T_3}{N_2 r_{\nu}} & 0 \\
 \frac{a f N_1 v_u c_{\theta } G_{\nu} T_2}{r_{\nu^c}} & -\frac{a f N_2 v_u Q_{\nu^c} T_1 T_3}{N_1 r_{\nu^c}} & \frac{a f^2 v_u c_{\theta } \left(N_1^2 G_{\nu}^2 Q_{\nu}-N_2^2 Q_{{\nu}^c} T_1^2\right) T_2 T_3}{N_1 N_2 G_{\nu} r_{\nu} r_{\nu^c}} &0\\
 0 & a N_2 v_uG_{\nu^c} & \frac{a f N_2 v_u c_{\theta } G_{\nu^c} T_1 T_2}{G_{\nu} r_{\nu}} & M_1 G_{\nu^c} Q_{\nu^c} T_3 \\
 a N_2 v_u G_{\nu} & 0 & \frac{a f v_u G_{\nu} Q_{\nu} T_3}{r_{\nu}} & M_1 N_2 c_{\theta } T_1 T_2 \\
 \frac{a f N_2 v_u c_{\theta } G_{\nu} T_1 T_2}{G_{\nu^c} r_{\nu^c}} & \frac{a f v_u G_{\nu^c} Q_{\nu^c} T_3}{r_{\nu^c}} & \frac{a f^2 v_u c_{\theta } \left(G_{\nu}^2 Q_{\nu}+G_{\nu^c}^2 Q_{\nu^c}\right) T_1 T_2 T_3}{G_{\nu} G_{\nu^c} r_{\nu} r_{\nu^c}} & -\frac{f M_1 \left(N_1^2 G_{\nu^c}^2+N_2^2 T_1^2\right)}{N_1^2 N_2G_{\nu^c} r_{\nu^c}} \\
 M_1 G_{\nu} Q_{\nu} T_3 & M_1 N_2 c_{\theta } T_1 T_2 & -\frac{f M_1 \left(N_1^2 G_{\nu}^2+N_2^2 T_1^2\right)}{N_1^2 N_2 G_{\nu} r_{\nu}} & 0 \\
 0 & \frac{M_2}{N_2} & 0 & 0 \\
 \frac{M_3}{G_{\nu}} & 0 & 0 & 0 \\
 0 & 0 & 0 & 0 \\
 0 & 0 & 0 & 0 \\
 0 & 0 & 0 & 0 \\
 0 & 0 & 0 & 0 \\
 0 & 0 & 0 & 0 \\
 0 & 0 & 0 & 0
\end{array}
\right),\nonumber
\end{eqnarray}

\begin{eqnarray}
\footnotesize\left(
\begin{array}{cccccc}
 0 & 0 & 0 & a N_1 v_u G_{\nu} s_{\theta } T_2 & \frac{a f N_1 v_u c_{\theta } G_{\nu} T_2}{r_{{\nu}^c}} & 0 \\
 0 & 0 & 0 & 0 & -\frac{a f N_2 v_u Q_{{\nu}^c} T_1 T_3}{N_1 r_{{\nu}^c}} & a N_2 v_u G_{{\nu}^c} \\
 0 & 0 & 0 & \frac{a f N_1 v_u G_{\nu} Q_{\nu} s_{\theta } T_2 T_3}{N_2 r_{\nu}} & \frac{a f^2 v_u c_{\theta } \left(N_1^2 G_{\nu}^2 Q_{\nu}-N_2^2 Q_{{\nu}^c} T_1^2\right) T_2 T_3}{N_1 N_2 G_{\nu} r_{\nu} r_{{\nu}^c}} & \frac{a f N_2 v_u c_{\theta } G_{{\nu}^c} T_1 T_2}{G_{\nu} r_{\nu}} \\
 0 & 0 & 0 & 0 & 0 & M_1 G_{{\nu}^c} Q_{{\nu}^c} T_3 \\
 0 & 0 & 0 & 0 & 0 & 0 \\
 0 & 0 & 0 & 0 & 0 & \frac{M_3}{G_{{\nu}^c}} \\
 0 & 0 & 0 & 0 & 0 & 0 \\
 0 & 0 & 0 & 0 & 0 & 0 \\
 0 & 0 & 0 & 0 & 0 & 0 \\
 0 & \frac{M_3}{G_{{\nu}^c}} & 0 & 0 & 0 & 0 \\
 \frac{M_2}{N_2} & 0 & 0 & 0 & 0 & 0 \\
 0 & 0 & 0 & 0 & 0 & 0 \\
 0 & 0 & 0 & 0 & 0 & 0 \\
 0 & 0 & 0 & 0 & 0 & 0 \\
 0 & 0 & 0 & 0 & 0 & 0 \\
 0 & 0 & v b_4 & 0 & 0 & 0 \\
 0 & 0 & 0 & -N_1 v b_4 & \frac{f N_1 v b_4 c_{\theta } s_{\theta } T_2^2}{r_{\nu}} & 0 \\
 0 & 0 & 0 & 0 & -\frac{f v b_4}{N_1 r_{\nu}} & 0 \\
 0 & 0 & 0 & 0 & 0 & 0 \\
 0 & 0 & 0 & 0 & 0 & 0 \\
 0 & 0 & 0 & 0 & 0 & 0
\end{array}\right),\nonumber
\end{eqnarray}
\begin{eqnarray}
\footnotesize\left(\begin{array}{cccccc}
 a N_2 v_u G_{\nu} & \frac{a f N_2 v_u c_{\theta } G_{\nu} T_1 T_2}{G_{\nu^c} r_{\nu^c}} & M_1 G_{\nu} Q_{\nu} T_3 & 0 & \frac{M_3}{G_{\nu}} & 0 \\
 0 & \frac{a f v_u G_{\nu^c} Q_{\nu^c} T_3}{r_{\nu^c}} & M_1 N_2 c_{\theta } T_1 T_2 & \frac{M_2}{N_2} & 0 & 0 \\
 \frac{a f v_u G_{\nu} Q_{\nu} T_3}{r_{\nu}} & \frac{a f^2 v_u c_{\theta } \left(G_{\nu}^2 Q_{\nu}+G_{\nu^c}^2 Q_{\nu^c}\right) T_1 T_2 T_3}{G_{\nu} G_{\nu^c} r_{\nu} r_{\nu^c}} & -\frac{f M_1 \left(N_1^2 G_{\nu}^2+N_2^2 T_1^2\right)}{N_1^2 N_2 G_{\nu} r_{\nu}} & 0 & 0 &0 \\
  M_1 N_2 c_{\theta } T_1 T_2 & -\frac{f M_1 \left(N_1^2 G_{\nu^c}^2+N_2^2 T_1^2\right)}{N_1^2 N_2 G_{\nu^c} r_{\nu^c}} & 0 & 0 & 0 & 0 \\
 \frac{M_2}{N_2} & 0 & 0 & 0 & 0 & 0 \\
 0 & 0 & 0 & 0 & 0 & 0 \\
 0 & 0 & 0 & 0 & 0 & v b_4 \\
 0 & 0 & 0 & 0 & 0 & 0 \\
 0 & 0 & 0 & 0 & 0 & 0 \\
 0 & 0 & 0 & 0 & 0 & 0 \\
 0 & 0 & 0 & 0 & 0 & 0 \\
 0 & 0 & 0 & 0 & 0 & 0 \\
 0 & 0 & 0 & 0 & 0 & 0 \\
 0 & 0 & 0 & 0 & 0 & 0 \\
 0 & 0 & 0 & 0 & 0 & 0 \\
 0 & 0 & 0 & 0 & 0 & 0 \\
 N_2 v b_4 s_{\theta } T_2 & \frac{f N_2 v b_4 c_{\theta } s_{\theta } T_1 T_2^2}{G_{\nu} r_{\nu}} & 0 & 0 & 0 & 0 \\
 N_2 v b_4 c_{\theta } T_2 & -\frac{f N_2 v b_4 T_1}{N_1^2 G_{\nu} r_{\nu}} & 0 & 0 & 0 & 0 \\
 0 & 0 & 0 & 0 & 0 & 0 \\
 0 & 0 & 0 & 0 & 0 & c \alpha_3 \\
 0 & 0 & 0 & 0 & 0 & c \alpha_2
\end{array}\right),\nonumber
\end{eqnarray}
\begin{eqnarray}
\footnotesize\left(
\begin{array}{ccccc}
 0 & 0 & 0 & 0 & 0 \\
 0 & 0 & 0 & 0 & 0 \\
 0 & 0 & 0 & 0 & 0 \\
 0 & 0 & 0 & 0 & 0 \\
 0 & 0 & 0 & 0 & 0 \\
 0 & 0 & 0 & 0 & 0 \\
 0 & 0 & 0 & 0 & 0 \\
 -N_1 v b_4 & 0 & 0 & 0 & 0 \\
 \frac{f N_1 v b_4 c_{\theta } s_{\theta } T_2^2}{r_{\nu}} & -\frac{f v b_4}{N_1 r_{\nu}} & 0 & 0 & 0 \\
 0 & 0 & 0 & 0 & 0 \\
 N_2 v b_4 s_{\theta } T_2 & N_2 v b_4 c_{\theta } T_2 & 0 & 0 & 0 \\
 \frac{f N_2 v b_4 c_{\theta } s_{\theta } T_1 T_2^2}{G_{\nu} r_{\nu}} & -\frac{f N_2 v b_4 T_1}{N_1^2 G_{\nu} r_{\nu}} & 0 & 0 & 0 \\
 0 & 0 & 0 & 0 & 0 \\
 0 & 0 & 0 & 0 & 0 \\
 0 & 0 & 0 & 0 & 0 \\
 0 & 0 & 0 & c \alpha_3 & c \alpha_2 \\
 0 & 0 & c \alpha_3 & 0 & c \alpha_1 \\
 0 & 0 & c \alpha_2 & c \alpha_1 & 0 \\
 c \alpha_3 & c \alpha_2 & m_1 & 0 & 0 \\
 0 & c \alpha_1 & 0 & m_1 & 0 \\
 c \alpha_1 & 0 & 0 & 0 & m_1
\end{array}
\right).
\end{eqnarray}
Here $v=\langle 1(\overline{16}_H)\rangle $. The light neutrino mass matrix is given by the seesaw formula as
fellows
\begin{eqnarray}
M_{\nu}&=& M_DM_R^{-1}M_D^T=\lambda\left(
\begin{array}{ccc}
 0 & 0 &0\\
 0 & \kappa &\eta \\
 0 & \eta&1
\end{array}\right),
\end{eqnarray}
where
\begin{eqnarray}
\lambda &=&\frac{\Lambda a^2 c^2 f^2 v_d^2 T_1^2 T_2^2
\left(\left(\alpha_1^2 +\alpha_3^2\right) Q_\nu^2 r_{\nu^c}^2
s_{\theta }^2+2 N_1^2 \alpha_2 \alpha_3 c_{\theta } Q_\nu r_{\nu^c}
s_{\theta } \left(\left(Q_\nu+Q_{\nu^c}\right)
r_\nu\right.\right.}{m_1
N_1^2 v^2 b_4^2 r_\nu^2 r_{\nu^c}^2}\nonumber\\
&+&\frac{\left.\left.Q_\nu r_{\nu^c} s_{\theta }^2
T_2^2\right)+N_1^4 \left(\alpha_1^2+\alpha_2^2\right) c_{\theta }^2
\left(\left(Q_\nu+Q_{\nu^c}\right) r_\nu+Q_\nu r_{\nu^c} s_{\theta
}^2 T_2^2\right){}^2\right) T_3^2}{m_1
N_1^2 v^2 b_4^2 r_\nu^2 r_{\nu^c}^2},\nonumber\\
\eta&=&\frac{N_1^2 Q_{\nu^c} r_\nu^2 s_{\theta } \left(\alpha_2
\alpha_3 Q_\nu r_{\nu^c} s_{\theta }+N_1^2 \left(\alpha_1^2+\alpha
_2^2\right) c_{\theta } \left(\left(Q_\nu+Q_{\nu^c}\right)
r_\nu+Q_\nu r_{\nu^c} s_{\theta
}^2 T_2^2\right)\right)}{f(A+B)},\nonumber\\
\kappa &=&\frac{N_1^4 \left(\alpha_1^2+\alpha_2^2\right) Q_{\nu^c}^2
r_\nu^4 s_{\theta }^2}{f^2(A+B)}.
\end{eqnarray}
Here the numerical values of $\alpha_1$, $\alpha_2$,  $\alpha_3$ and
$\lambda$ are given in section 2, and we have defined
\begin{eqnarray}
A&=&\left(\alpha_1^2+\alpha_3^2\right) Q_e^2 r_{e^c}^2 s_{\theta
}^2+2 N_1^2 \alpha_2 \alpha_3 c_{\theta } Q_e r_{e^c} s_{\theta }
\left(\left(Q_e+Q_{e^c}\right) r_e+Q_e r_{e^c} s_{\theta }^2
T_2^2\right), \nonumber\\
B&=&N_1^4 \left(\alpha_1^2+\alpha_2^2\right) c_{\theta }^2
\left(\left(Q_e+Q_{e^c}\right) r_e+Q_e r_{e^c} s_{\theta }^2
T_2^2\right).\nonumber
\end{eqnarray}
\section{RGEs from the scale $M^*$ to the GUT scale}
Neglecting all the couplings in the superpotential $W_{vector}$,
since they do not contribute to the top Yukawa coupling, we present
only the RGEs that are needed to find the parameter $\sigma$ at the
GUT scale. The one-loop RGE's of the unified gauge ($g_G$) coupling,
the couplings appearing in $W_{spinor}$, and the trilinear soft
terms associated with $W_{spinor}$  between the scale $M^*$ and GUT
scale are
\begin{eqnarray}
16\pi^2\frac {dg_{G}}{dt}&=&19g_{G}^3,\\
16\pi^2\frac {db_1}{dt}&=&b_1(20b_1^2+b_2^2-45g_{G}^2),\\
16\pi^2\frac {db_2}{dt}&=&b_2(20b_2^2+b_1^2-45g_{G}^2),\\
16\pi^2\frac {da}{dt}&=&a(18a^2-\frac{63}{2}g_{G}^2),\\
16\pi^2\frac {d\tilde{b}_1}{dt}&=&2(20b_1^2\tilde{b}_1+b_2^2\tilde{b}_2+45g_{G}^2M_{\lambda}),\\
16\pi^2\frac {d\tilde{b}_2}{dt}&=&2(20b_2^2\tilde{b}_2+b_1^2\tilde{b}_1+45g_{G}^2M_{\lambda}),\\
16\pi^2\frac {d\tilde{a}}{dt}&=&28\tilde{a}a^2+63g_{G}^2M_{\lambda}.
\end{eqnarray}
The RGE's soft mass terms for the fields appearing in $W_{spinor}$
are given below:
\begin{eqnarray}
16\pi^2\frac {d\tilde{m}^2_{\psi_i}}{dt}&=&2b_1^2(\tilde{m}^2_{\psi_i}+\tilde{m}^2_{\overline{\chi}_1}+\tilde{m}^2_{1_{H_i}}+\tilde{b}^2_1)\nonumber\\
&+&2b_2^2(\tilde{m}^2_{\psi_i}+\tilde{m}^2_{\overline{\chi}_2}+\tilde{m}^2_{1'_{H_i}}+\tilde{b}^2_2)-45g_{G}^2M_{\lambda}^2,\\
16\pi^2\frac {d\tilde{m}^2_{\overline{\chi}_1}}{dt}&=&6b_1^2(\tilde{m}^2_{\psi_i}+\tilde{m}^2_{\overline{\chi}_1}+\tilde{m}^2_{1_{H_i}}+\tilde{b}^2_1)-45g_{G}^2M_{\lambda}^2,\\
16\pi^2\frac {d\tilde{m}^2_{\overline{\chi}_2}}{dt}&=&6b_2^2(\tilde{m}^2_{\psi_i}+\tilde{m}^2_{\overline{\chi}_2}+\tilde{m}^2_{1'_{H_i}}+\tilde{b}^2_2)-45g_{G}^2M_{\lambda}^2,\\
16\pi^2\frac {d\tilde{m}^2_{1_{H_i}}}{dt}&=&32b_1^2(\tilde{m}^2_{\psi_i}+\tilde{m}^2_{\overline{\chi}_1}+\tilde{m}^2_{1_{H_i}}+\tilde{b}^2_1),\\
16\pi^2\frac {d\tilde{m}^2_{1'_{H_i}}}{dt}&=&32b_2^2(\tilde{m}^2_{\psi_i}+\tilde{m}^2_{\overline{\chi}_2}+\tilde{m}^2_{1'_{H_i}}+\tilde{b}^2_2),\\
16\pi^2\frac {d\tilde{m}^2_{\chi_{2,3}}}{dt}&=&10a^2(\tilde{m}^2_{\chi_2}+\tilde{m}^2_{\chi_3}+\tilde{m}^2_{10_{H}}+\tilde{a}^2)-45g_{G}^2M_{\lambda}^2,\label{p3r}\\
16\pi^2\frac
{d\tilde{m}^2_{10_H}}{dt}&=&16a^2(\tilde{m}^2_{\chi_2}+\tilde{m}^2_{\chi_3}+\tilde{m}^2_{10_{H}}+\tilde{a}^2)-36g_{G}^2M_{\lambda}^2.
\end{eqnarray}
Here $\tilde{m}^2_{1_{H_i}}$, $\tilde{m}^2_{1'_{H_i}}$ and
$\tilde{m}^2_{10_{H}}$ are the quadratic soft masses for the Higgs
superfields appearing in $W_{spin}$ defined in Eq.(\ref{p32}) and
the quadratic soft masses $\tilde{m}^2_{\psi_i}$,
$\tilde{m}^2_{\overline{\chi}_{1,2}}$, and
$\tilde{m}^2_{\chi_{1,2}}$ are defined in Eq.(\ref{p35}).

{}

\end{document}